# Sensing Spin Wave Excitations by Spin Defects in Few-Layer Thick Hexagonal Boron Nitride

Jingcheng Zhou,[1] Hanyi Lu,[2] Di Chen,[3,4] Mengqi Huang,[1] Gerald Q. Yan,[2] Faris Al-matouq,[1] Jiu Chang,[1] Dziga Djugba,[1] Zhigang Jiang,[1] Hailong Wang,[1,*] Chunhui Rita Du[1,2,*]

[1]School of Physics, Georgia Institute of Technology, Atlanta, Georgia 30332
[2]Department of Physics, University of California, San Diego, La Jolla, California 92093
[3]Department of Physics, University of Houston, Houston, Texas 77204
[4]Texas Center for Superconductivity, University of Houston, Houston, Texas 77204

[*]Corresponding authors: hwang3021@gatech.edu; cdu71@gatech.edu

**Abstract**: Optically active spin defects in wide band-gap semiconductors serve as a local sensor of multiple degrees of freedom in a variety of "hard" and "soft" condensed matter systems. Taking advantage of the recent progress on quantum sensing using van der Waals (vdW) quantum materials, here we report direct measurements of spin waves excited in magnetic insulator $Y_3Fe_5O_{12}$ (YIG) by boron vacancy $V_B^-$ spin defects contained in few-layer thick hexagonal boron nitride nanoflakes. We show that the ferromagnetic resonance and parametric spin excitations can be effectively detected by $V_B^-$ spin defects under various experimental conditions through optically detected magnetic resonance measurements. The off-resonant dipole interaction between YIG magnons and $V_B^-$ spin defects is mediated by multi-magnon scattering processes, which may find relevant applications in a range of emerging quantum sensing, computing, and metrology technologies. Our results also highlight the opportunities offered by quantum spin defects in layered two-dimensional vdW materials for investigating local spin dynamic behaviors in magnetic solid-state matters.

**Introduction**

Nanoscale control and detection of spin waves, quanta of spin angular momentum in magnetic systems also referred to as magnons, constitutes a key challenge for the development of modern solid-state spintronic technologies (*1*, *2*). Optically active spin defects in wide band-gap semiconductors are finding increased applications in this field due to their competitive magnetic sensitivity, spatial resolution, and remarkable functionality over a broad range of experimental conditions (*3–7*). Over the past decade, nitrogen-vacancy (NV) centers, one of the most prominent candidates of this class of defects, have been applied as effective sensors of local spin transport and dynamic behaviors in various condensed matter systems (*8*). Notable examples include off-resonant NV detection of magnon excitations by nanodiamonds (*9–11*), nanoscale measurements of spin chemical potential in a magnetic insulator (*12*), probing spin torque driven auto oscillations (*13*, *14*), magnetic resonance imaging of spin-wave propagation (*15*, *16*), and others (*8*, *17*). More recently, emergent color centers hosted in layered van der Waals (vdW) crystals such as transition metal dichalcogenides and hexagonal boron nitride (hBN) became new contenders in this field (*18*, *19*). In comparison with NV centers embedded in highly rigid, three-dimensional (3D) diamond, spin defects contained in exfoliable, two-dimensional (2D) materials exhibit improved compatibility with nanodevice integration, providing an attractive platform for implementing ultra-sensitive quantum metrology measurements by exploiting the atomic length scale proximity between the spin sensors and objects of interest (*20–23*). In the current state of the art, the local sensing of electrical, magnetic, and thermal flux arising from solid-state materials using spin defects in hBN has been experimentally demonstrated in both confocal and wide-field optical microscopy configurations (*5*, *21*, *22*, *24*), and the development of transformative approaches to revolutionize the current quantum technologies are under way (*25–29*).

Despite the much progress over the past years, to date, experimental demonstration of practical quantum sensing using few-layer thick hBN in a real, "non-sterile" material environment remains a formidable challenge. In this work, we present our recent efforts along this direction. Specifically, we report quantum sensing of magnons with variable wavevectors in a magnetic insulator $Y_3Fe_5O_{12}$ (YIG) (*30*) by boron vacancy $V_B^-$ spin defects in hBN flakes with a thickness down to ~2 nanometers. Using optically detected magnetic resonance (ODMR) measurements (*5*, *18*), we show that ferromagnetic resonance (FMR) and parametric pumping (*31*, *32*) driven YIG spin waves can dynamically interact with $V_B^-$ spin defects to induce "three-level" spin transitions with associated variations in the $V_B^-$ photoluminescence (PL). The YIG spin wave spectrum measured under a broad range of experimental conditions can be well interpreted by considering the sample's magnon band structures and intrinsic magnetic properties. The sensitivity length scale and magnetic field resolution of the presented hBN-based quantum sensing platform are primarily determined by the distance between the $V_B^-$ spin defects and the magnetic sample (*21*, *22*), which can readily reach the atomic length scale due to the established vdW proximity. The demonstrated coupling between spin defects in 2D vdW crystals and magnons carried by a magnetic insulator brings opportunities for developing hybrid solid-state electronics for next-generation quantum information and technological applications (*33*).

**Results**

In the current study, we utilized two independent methods including Helium ion (He$^+$) implantation and thermal neutron irradiations to create boron vacancy $V_B^-$ spin defects in hBN (See Materials and Methods for details). Figure 1A presents our measurement platform, where an exfoliated hBN flake containing $V_B^-$ spin defects is mechanically transferred onto an Au microwave stripline patterned on top of a 100-nm-thick YIG thin film. Note that the $V_B^-$ spins are orientated along the out-of-plane (OOP) direction of the hBN sample (*18*) and microwave signals delivered into the on-chip Au stripline are used to provide local control of the $V_B^-$ spin states and YIG magnon excitations. The $V_B^-$ fluorescence is measured using an avalanche photo diode in a confocal microscope. An optical microscope image shown in Fig. 1B provides an overview of a prepared YIG-hBN device (sample A) for ODMR measurements. The thickness of the surveyed hBN flake area is determined to be ~6 nm from layer dependent optical contrast (Fig. 1C), which is confirmed by atomic force microscopy measurements shown in Fig. 1D.

Figure 1E presents the energy level diagram of the ground state (GS), excited state (ES), and metastable state (MS) of a $V_B^-$ spin defect. The negatively charged $V_B^-$ state has an $S = 1$ electron spin and serves as a three-level system (*18*). Fluctuating magnetic fields at the corresponding electron spin resonance (ESR) frequencies will induce $V_B^-$ spin transitions between the $m_s = 0$ and $m_s = \pm 1$ states (*21*). The $m_s = \pm 1$ $V_B^-$ spin states are more likely to be trapped by a non-radiative pathway through the metastable state before decaying back to the $m_s = 0$ ground state, generating reduced PL in the red wavelength range (*5*, *18*, *34*). The optically accessible $V_B^-$ spin states and their dipole interaction with the local magnetic field environment provide a suitable way to investigate spin wave excitations in a proximal magnetic material. In this work, we utilized both FMR and nonlinear parametric pumping to excite magnons with a broad range of wavevectors in the ferrimagnetic insulator YIG. Figure 1F shows the underlying mechanism of these two driving methods. For FMR excitation, the external microwave magnetic field $B_{mw}$ at a frequency $f_{mw}$ perpendicularly applied to the static YIG magnetization *M* drives quasi-uniform spin wave modes with a wavevector $k \approx 0$ at the same frequency of $f_{mw}$ (*32*, *35*). In contrast, parametric excitation features elliptically shaped precession of the YIG magnetization, where a sufficiently large $B_{mw}$ parallel with *M* excites exchange magnons ($k \geq 10$ rad/μm) at half the frequency ($f_{mw}/2$) of the microwave magnetic field (*32*, *35*, *36*). In a more classical spin dynamics picture, FMR refers to direct excitation of precessional motion of the transverse magnetization component while parametric pumping characterizes the microwave-driven oscillations of the longitudinal magnetization component (*32*).

We now present ODMR measurements of microwave driven YIG spin waves by $V_B^-$ spin defects. Figure 2A shows the schematic configuration of the static magnetic field $B_{ext}$, microwave magnetic field $B_{mw}$, and YIG magnetization *M* in our measurements. The vector direction of $B_{ext}$ is characterized by the spherical angles $\theta$ and $\varphi$. The $B_{mw}$ generated by the Au stripline is parallel to the *x*-axis direction and the YIG magnetization is partially tilted away from the sample plane due to the shape magnetic anisotropy. The current study focuses on a 100-nm-thick YIG film, whose magnon band structure is shown in Fig. 2B (see section S1 for details). In the small wavevector regime of $k \leq 1$ rad/μm, the YIG spin wave energy is dominated by dipole interaction (*36*). Magnetic dipole fields generated by the dynamic magnetization along the thickness direction of the YIG sample increase and decrease the energy of the surface spin wave mode and the back volume spin wave mode relative to the FMR frequency $f_{FMR}$, respectively (*30*). In the large wavevector regime ($k \geq 10$ rad/μm), the exchange interaction dominates resulting in a parabolic increase of the spin wave energy. Under external microwave driving, FMR and parametric pumping excited magnons at frequencies $f_{FMR}$ and $f_{mw}/2$ interact with the YIG thermal magnon bath via multi-magnon scattering processes, establishing a new magnetic thermal equilibrium state with enhanced spin densities at the $V_B^-$ ESR frequencies $f_\pm$ (*37*). Here, $f_\pm$ represent the $V_B^-$ ESR frequencies corresponding to the spin transition between the $m_s = 0$ and $m_s = \pm 1$ states. The increased magnetic fluctuations at frequencies $f_\pm$ will induce spin relaxation of proximal $V_B^-$ spin defects and variations of the emitted PL, which can be optically detected.

We applied continuous green laser and microwave signals to perform the ODMR measurements (20). Figure 2C shows normalized PL of $V_B^-$ spin defects measured as a function of the microwave frequency $f_{mw}$ and external magnetic field $B_{ext}$. The polar and azimuthal angles of $B_{ext}$ are 16 and 39 degrees in this measurement and the input microwave power is 33 dBm. Under this field configuration, the microwave magnetic field $B_{mw}$ has both significant longitudinal and transverse components relative to the static YIG magnetization, allowing for the co-existence of FMR and parametric spin excitations. The two "straight" lines from ~3.5 GHz result from the expected, ESR-induced decrease of $V_B^-$ fluorescence. The measured $V_B^-$ PL is also reduced when $f_{mw}$ matches the calculated FMR frequency $f_{FMR}$ of the 100-nm-thick YIG film as shown by the lower branch of the white dashed lines (see section S1 for details). Notably, spin excitations also emerge in the high frequency regime with a threshold value of twice of the magnon band minimum $f_{min}$, exhibiting the key feature of parametric spin excitations (36, 37). Figure 2D shows a linecut at $B_{ext}$ = 204 Oe of the ODMR map measured over a broad range of input microwave powers from 20 dBm to 34 dBm. The $V_B^-$ PL intensity decreases at the corresponding frequencies of $f_{FMR}$, $2f_{min}$, $f_-$, and $f_+$. It is worth mentioning that the emergence of parametric pumping at frequencies above $2f_{min}$ exhibits a threshold microwave power of ~27 dBm. We also perform the ODMR measurements under different external magnetic field conditions. When $B_{ext}$ is 42 degrees tilted away from the OOP direction with the same azimuthal angle, the reduced demagnetizing field contributes to an increased YIG magnon energy due to the enhanced internal field experienced by the YIG sample (38). In this case, the measured FMR and parametric spin excitation frequencies increase accordingly as shown in Fig. 2E. Note that the $V_B^-$ ESR lines are curved due to the notable misalignment between $B_{ext}$ and the $V_B^-$ spin direction (18).

Next, we systematically modulate the static magnetic field direction to investigate variations of the FMR and parametric spin wave excitation efficiency in the YIG sample. Figure 3A presents a normalized PL map of $V_B^-$ spin defects measured as a function of the microwave frequency $f_{mw}$ and azimuthal angle $\varphi$ of $B_{ext}$. The magnitude and polar angle $\theta$ of $B_{ext}$ are fixed to be 150 Oe and 16 degrees in this measurement, and the input microwave power is 33 dBm. The FMR and parametric spin waves exhibit clear angular dependence on $\varphi$ and reach a maximum magnitude under different field conditions. To further highlight this point, Fig. 3B plots a series of linecuts for varying in-plane azimuthal angle $\varphi$. When $\varphi = 0$ degree, the local microwave magnetic field $B_{mw}$ is mostly aligned parallel with the static YIG magnetization $M$, resulting in the highest parametric pumping driving efficiency. In contrast, the conventional FMR spin excitations are largely suppressed under this field geometry due to the minimal transverse microwave magnetic field component. The static YIG magnetization $M$ rotates accordingly around the z-axis when $\varphi$ varies due to the negligibly small in-plane magnetic anisotropy. When $\varphi = 90$ degrees, $M$ is mostly perpendicular with $B_{mw}$, leading to a maximum FMR excitation efficiency while the signature of parametric pumping virtually disappears. Figures 3C and 3D show two PL maps of $V_B^-$ spin defects measured at $\varphi = 0$ and 90 degrees, respectively, providing an overview of the observed YIG spin wave spectrum. One can see that the FMR and parametric spin excitations emerge and vanish under the corresponding field geometries, in agreement with the picture discussed above. Note that multiple few-layer thick hBN samples have been evaluated in our work to ensure the reproducibility of the presented results. In the Supplementary Materials Section S2, we present extended ODMR results on a 2-nm-thick hBN flake (sample B, see section S2 for details), demonstrating the potential to push the hBN thickness to the atomical scale for quantum sensing study. Here, we further comment that the advantage of placing atomically thin hBN in vdW proximity to desired positions of a magnetic material allows for sensitive detection of high-wavevector spin waves. Based on the magnon dispersion curves calculated when the wavevector is parallel and perpendicular to the in-plane projection of the YIG magnetization(*38*, *39*), it is estimated that the wavevector of YIG spin excitations measured in the current work lies in the range of $3 \times 10^7$ m$^{-1}$ ≤ $k$ ≤ $6 \times 10^7$ m$^{-1}$ (see section S3 for details), which is not accessible in our previous work using NV centers (*37*).

Lastly, we present extended ODMR measurements on a separate 76-nm-thick hBN flake (sample C, see section S4 for details) to demonstrate the quantum sensing functionality of $V_B^-$ spin defects over a broad range of temperature. Figures 4A and 4B show two ODMR maps recorded at 200 K and 280 K. The external magnetic field $B_{ext}$ is applied along the direction $\theta$ = 16 degrees and $\varphi$ = 15 degrees in these measurements. Both the FMR and parametric spin wave excitations are observed at the corresponding resonant frequencies (see section S4 for details). The moderate enhancement of $f_{FMR}$ and $2f_{min}$ in the low temperature regime is attributed to the increased static YIG magnetization (see section S1 for details). Figure 4C summarizes the magnetic field dependence of the FMR frequency $f_{FMR}$ measured in a temperature range from 100 K to 300 K, in agreement with the theoretically calculated YIG spin wave dispersion curves (see section S1 for details) (*38*). Here, we briefly comment on the temperature dependent variations of the optical contrast of the observed YIG spin waves. Qualitatively, the optical contrast of the spin waves reflects the number of YIG magnons "pumped" to the $V_B^-$ ESR frequencies by FMR or parametric spin excitations. From a statistical viewpoint, the occupation of YIG magnons, bosonic-type quasiparticles, is expected to follow the classical Bose-Einstein distribution in a thermal equilibrum state (*12*, *40*). Under our experimental conditions where the thermal energy is much larger than the magnon energy, the YIG magnon density at a given frequency is proportional to $\sim k_B T$, where $k_B$ is the Boltzmann constant and $T$ is temperature (*12*, *40*). Thus, the YIG magnon density will decrease at low temperatures, resulting in reduced optical contrast of spin waves as shown in Fig. 4D. In fact, this point is also corroborated by $V_B^-$-based quantum sensing study of thermally induced YIG magnon noise. As a qubit-based magnetometer, $V_B^-$ spin defects provide the possibility of probing noncoherent magnetic fields that are challenging to access by conventional magnetometry methods. Fluctuating magnetic fields generated by YIG magnons at the ESR frequencies will induce m$_s$ = 0 ↔ ±1 transitions of $V_B^-$ spin defects, which can be optically detected by spin relaxometry measurements (*12*, *21*). We observe that the measured YIG magnon noise qualitatively decreases with decreasing temperature from 250 K to 20 K, indicating a reduced YIG magnon density in the low temperature regime (see section S5 for details). It is also instructive to note that the YIG FMR and parametric pumping efficiency gets dramatically decreased below 200 K possibly due to a combined effect of variations of magnon density (*12*), Gilbert damping (*41*) and the efficiency of pumping incoherent magnon gas through coherent magnetization dynamics (*12*, *42*). A comprehensive understanding of these effects calls for future systematic studies.

**Discussion**

In summary, we have demonstrated $V_B^-$ spin defects contained in few-layer thick hBN flakes to be a versatile local sensor of proximal spin wave excitations under a broad range of experimental conditions (see section S6 for details). Robust FMR and nonlinear parametric spin excitations are observed in a 100-nm-thick YIG film by $V_B^-$-based ODMR measurements. Detailed sample information such as the local static magnetization, spin fluctuations, exchange stiffness, and magnon band minimum can be extracted from the presented quantum sensing measurements, highlighting the capabilities of $V_B^-$ spin defects in accessing local spin dynamic behaviors in magnetic materials. We expect that the improved solid-state scalability and compatibility with nanodevice fabrication will establish hBN as an excellent candidate for implementing next-generation quantum sensing innovations. The off-resonant dipole interaction between magnons and $V_B^-$ spin defects may also find potential applications in developing hybrid quantum systems consisting of layered 2D vdW crystals and functional spintronic devices for emerging quantum information and science applications (*33*).

**Materials and Methods**
**Sample fabrication**

Two independent spin defect generation methods are employed in the current study. The first one is thermal neutron irradiation of monoisotopic hBN crystals with a dose of ~$2.6 \times 10^{16}$ neutrons·cm$^{-2}$. The $^{10}$B-enriched monoisotopic hBN crystals are synthesized using a metal (Ni-Cr) flux method (*43*). The starting materials include 48 wt.% Ni (Alfa Aesar, 99.70%), 48 wt.% Cr (Alfa Aesar, 99.99%), and 4 wt.% $^{10}$B (Ceradyne 3M, 96.85%) powders. The source of elemental nitrogen is flowing N$_2$ gas (ultra-high purity, 125 sccm), and 5 sccm H$_2$ gas is used to minimize impurities originating from oxygen and carbon. Samples A (6-nm-thick hBN), B (2-nm-thick hBN), and D (75-nm-thick hBN) are mechanically exfoliated from neutron irradiated $^{10}$B-enriched monoisotopic hBN crystals and then transferred onto the YIG film for quantum sensing measurements. Samples A and B are utilized for ODMR measurements and sample D is for $V_B^-$ spin relaxometry measurements of YIG magnon noise.

The second method is Helium ion (He$^+$) implantation of hBN flakes pre-exfoliated from standard hBN crystals commercially available from HQ Graphene. Boron vacancy $V_B^-$ spin defects in hBN sample C and sample E are created in this way with a Helium ion energy of 5 keV and dose of $5 \times 10^{13}$ cm$^{-2}$ using 200 kV NEC ion implantation at room temperature. The thickness of the hBN sample C is ~76 nm and the average ion implantation depth of the $V_B^-$ spin defects is estimated to be ~60 nm by Stopping and Range of Ions in Matter (SRIM) simulations. Relevant optical microscopy images and thickness calibrations of the prepared hBN samples are presented in Fig. 1 and Supplementary Materials. The 100-nm-thick YIG films grown on (111)-oriented Gd$_3$Ga$_5$O$_{12}$ substrates were commercially available from the company Matesy GmbH.

**ODMR measurements**

The presented quantum sensing measurements were performed by a custom-designed confocal microscope with samples positioned inside a closed-loop optical cryostat. We applied continuous green laser and microwave signals to perform ODMR measurements. The PL of $V_B^-$ spin defects was detected by an avalanche photo diode and the external magnetic field was generated by a cylindrical NdFeB permanent magnet attached to a motorized translation stage inside the optical cryostat.

**References**:

1. A. Barman, G. Gubbiotti, S. Ladak, A. O. Adeyeye, M. Krawczyk, J. Grafe, C. Adelmann, S. Cotofana, A. Naeemi, V. I. Vasyuchka, B. Hillebrands, S. A. Nikitov, H. Yu, D. Grundler, A. V. Sadovnikov, A. A. Grachev, S. E. Sheshukova, J. Y. Duquesne, M. Marangolo, G. Csaba, W. Porod, V. E. Demidov, S. Urazhdin, S. O. Demokritov, E. Albisetti, D. Petti, R. Bertacco, H. Schultheiss, V. V. Kruglyak, V. D. Poimanov, S. Sahoo, J. Sinha, H. Yang, M. Münzenberg, T. Moriyama, S. Mizukami, P. Landeros, R. A. Gallardo, G. Carlotti, J. V. Kim, R. L. Stamps, R. E. Camley, B. Rana, Y. Otani, W. Yu, T. Yu, G. E. W. Bauer, C. Back, G. S. Uhrig, O. V. Dobrovolskiy, B. Budinska, H. Qin, S. van Dijken, A. V. Chumak, A. Khitun, D. E. Nikonov, I. A. Young, B. W. Zingsem, M. Winklhofer, The 2021 magnonics roadmap. *J. Phys. Condens. Matter* **33**, 413001 (2021).
2. A. V. Chumak, V. I. Vasyuchka, A. A. Serga, B. Hillebrands, Magnon spintronics. *Nat. Phys.* **11**, 453–461 (2015).
3. C. L. Degen, F. Reinhard, P. Cappellaro, Quantum sensing. *Rev. Mod. Phys.* **89**, 035002 (2017).
4. L. Rondin, J. P. Tetienne, T. Hingant, J. F. Roch, P. Maletinsky, V. Jacques, Magnetometry with nitrogen-vacancy defects in diamond. *Rep. Prog. Phys.* **77**, 056503 (2014).
5. S. Vaidya, X. Gao, S. Dikshit, I. Aharonovich, T. Li, Quantum sensing and imaging with spin defects in hexagonal boron nitride. *Adv. Phys. X* **8**, 2206049 (2023).
6. W. F. Koehl, B. B. Buckley, F. J. Heremans, G. Calusine, D. D. Awschalom, Room temperature coherent control of defect spin qubits in silicon carbide. *Nature* **479**, 84–87 (2011).
7. A. Sipahigil, R. E. Evans, D. D. Sukachev, M. J. Burek, J. Borregaard, M. K. Bhaskar, C. T. Nguyen, J. L. Pacheco, H. A. Atikian, C. Meuwly, R. M. Camacho, F. Jelezko, E. Bielejec, H. Park, M. Lončar, M. D. Lukin, An integrated diamond nanophotonics platform for quantum-optical networks. *Science* **354**, 847–850 (2016).
8. F. Casola, T. van der Sar, A. Yacoby, Probing condensed matter physics with magnetometry based on nitrogen-vacancy centres in diamond. *Nat. Rev. Mater.* **3**, 17088 (2018).
9. C. S. Wolfe, V. P. Bhallamudi, H. L. Wang, C. H. Du, S. Manuilov, R. M. Teeling-Smith, A. J. Berger, R. Adur, F. Y. Yang, P. C. Hammel, Off-resonant manipulation of spins in diamond via precessing magnetization of a proximal ferromagnet. *Phys. Rev. B* **89**, 180406 (2014).
10. B. A. McCullian, A. M. Thabt, B. A. Gray, A. L. Melendez, M. S. Wolf, V. L. Safonov, D. V. Pelekhov, V. P. Bhallamudi, M. R. Page, P. C. Hammel, Broadband multi-magnon relaxometry using a quantum spin sensor for high frequency ferromagnetic dynamics sensing. *Nat. Commun.* **11**, 5229 (2020).
11. J. Rable, B. Piazza, J. Dwivedi, N. Samarth, Local ferromagnetic resonance measurements of mesoscopically patterned ferromagnets using deterministically placed nanodiamonds. *Phys. Rev. Appl.* **18**, 064004 (2022).
12. C. Du, T. van der Sar, T. X. Zhou, P. Upadhyaya, F. Casola, H. Zhang, M. C. Onbasli, C. A. Ross, R. L. Walsworth, Y. Tserkovnyak, A. Yacoby, Control and local measurement of the spin chemical potential in a magnetic insulator. *Science* **357**, 195–198 (2017).
13. H. Zhang, M. J. H. Ku, F. Casola, C. H. R. Du, T. van der Sar, M. C. Onbasli, C. A. Ross, Y. Tserkovnyak, A. Yacoby, R. L. Walsworth, Spin-torque oscillation in a magnetic insulator probed by a single-spin sensor. *Phys. Rev. B.* **102**, 024404 (2020).

14. A. Solyom, M. Caouette-Mansour, B. Ruffolo, P. Braganca, L. Childress, J. C. Sankey, Single-spin spectroscopy of spontaneous and phase-locked spin-torque-oscillator dynamics. *Phys. Rev. Appl.* **20**, 54055 (2023).
15. I. Bertelli, J. J. Carmiggelt, T. Yu, B. G. Simon, C. C. Pothoven, G. E. W. Bauer, Y. M. Blanter, J. Aarts, T. van der Sar, Magnetic resonance imaging of spin-wave transport and interference in a magnetic insulator. *Sci. Adv.* **6**, eabd3556 (2020).
16. P. Andrich, C. F. de las Casas, X. Liu, H. L. Bretscher, J. R. Berman, F. J. Heremans, P. F. Nealey, D. D. Awschalom, Long-range spin wave mediated control of defect qubits in nanodiamonds. *npj Quantum Inf.* **3**, 28 (2017).
17. A. Finco, V. Jacques, Single spin magnetometry and relaxometry applied to antiferromagnetic materials. *APL Mater.* **11**, 100901 (2023).
18. A. Gottscholl, M. Kianinia, V. Soltamov, S. Orlinskii, G. Mamin, C. Bradac, C. Kasper, K. Krambrock, A. Sperlich, M. Toth, I. Aharonovich, V. Dyakonov, Initialization and read-out of intrinsic spin defects in a van der Waals crystal at room temperature. *Nat. Mater.* **19**, 540–545 (2020).
19. Y. Lee, Y. Hu, X. Lang, D. Kim, K. Li, Y. Ping, K. M. C. Fu, K. Cho, Spin-defect qubits in two-dimensional transition metal dichalcogenides operating at telecom wavelengths. *Nat. Commun.* **13**, 7501 (2022).
20. A. Gottscholl, M. Diez, V. Soltamov, C. Kasper, D. Krauße, A. Sperlich, M. Kianinia, C. Bradac, I. Aharonovich, V. Dyakonov, Spin defects in hBN as promising temperature, pressure and magnetic field quantum sensors *Nat. Commun.* **12**, 4480 (2021).
21. M. Huang, J. Zhou, D. Chen, H. Lu, N. J. McLaughlin, S. Li, M. Alghamdi, D. Djugba, J. Shi, H. Wang, C. R. Du, Wide field imaging of van der Waals ferromagnet $Fe_3GeTe_2$ by spin defects in hexagonal boron nitride. *Nat. Commun.* **13**, 5369 (2022).
22. A. J. Healey, S. C. Scholten, T. Yang, J. A. Scott, G. J. Abrahams, I. O. Robertson, X. F. Hou, Y. F. Guo, S. Rahman, Y. Lu, M. Kianinia, I. Aharonovich, J.-P. Tetienne, Quantum microscopy with van der Waals heterostructures. *Nat. Phys.* **19**, 87–91 (2022).
23. X. Lyu, Q. Tan, L. Wu, C. Zhang, Z. Zhang, Z. Mu, J. Zúñiga-Pérez, H. Cai, W. Gao, Strain quantum sensing with spin defects in hexagonal boron nitride. *Nano Lett.* **22**, 6553–6559 (2022).
24. P. Kumar, F. Fabre, A. Durand, T. Clua-Provost, J. Li, J. H. Edgar, N. Rougemaille, J. Coraux, X. Marie, P. Renucci, C. Robert, I. Robert-Philip, B. Gil, G. Cassabois, A. Finco, V. Jacques, Magnetic imaging with spin defects in hexagonal boron nitride. *Phys. Rev. Appl.* **18**, L061002 (2022).
25. X. Gao, S. Vaidya, K. Li, P. Ju, B. Jiang, Z. Xu, A. E. L. Allcca, K. Shen, T. Taniguchi, K. Watanabe, S. A. Bhave, Y. P. Chen, Y. Ping, T. Li, Nuclear spin polarization and control in hexagonal boron nitride. *Nat. Mater.* **21**, 1024–1028 (2022).
26. R. Gong, G. He, X. Gao, P. Ju, Z. Liu, B. Ye, E. A. Henriksen, T. Li, C. Zu, Coherent dynamics of strongly interacting electronic spin defects in hexagonal boron nitride. *Nat. Commun.* **14**, 3299 (2023).
27. N. Mathur, A. Mukherjee, X. Gao, J. Luo, B. A. McCullian, T. Li, A. N. Vamivakas, G. D. Fuchs, Excited-state spin-resonance spectroscopy of $V_B^-$ defect centers in hexagonal boron nitride. *Nat. Commun.* **13**, 3233 (2022).
28. A. Durand, T. Clua-Provost, F. Fabre, P. Kumar, J. Li, J. H. Edgar, P. Udvarhelyi, A. Gali, X. Marie, C. Robert, J. M. Gérard, B. Gil, G. Cassabois, V. Jacques, Optically active spin defects in few-layer thick hexagonal boron nitride. *Phys. Rev. Lett.* **131**, 116902 (2023).

29. X. Gao, S. Vaidya, S. Dikshit, P. Ju, K. Shen, Y. Jin, S. Zhang, T. Li, Nanotube spin defects for omnidirectional magnetic field sensing. arXiv:2310.02709 [cond-mat.mes-hall] (2023).
30. A. A. Serga, A. V. Chumak, B. Hillebrands, YIG magnonics. *J. Phys. D* **43**, 264002 (2010).
31. C. Hahn, G. De Loubens, M. Viret, O. Klein, V. V. Naletov, J. Ben Youssef, Detection of microwave spin pumping using the inverse spin hall effect. *Phys. Rev. Lett.* **111**, 217204 (2013).
32. F. Guo, L. M. Belova, R. D. McMichael, Parametric pumping of precession modes in ferromagnetic nanodisks. *Phys. Rev. B* **89**, 104422 (2014).
33. D. D. Awschalom, C. R. Du, R. He, F. Joseph Heremans, A. Hoffmann, J. Hou, H. Kurebayashi, Y. Li, L. Liu, V. Novosad, J. Sklenar, S. E. Sullivan, D. Sun, H. Tang, V. Tyberkevych, C. Trevillian, A. W. Tsen, L. R. Weiss, W. Zhang, X. Zhang, L. Zhao, C. H. W. Zollitsch, Quantum engineering with hybrid magnonic systems and materials. *IEEE Trans. Quantum Eng.* **2**, 1-36 (2021).
34. C. Qian, V. Villafañe, M. Schalk, G. V Astakhov, U. Kentsch, M. Helm, P. Soubelet, N. P. Wilson, R. Rizzato, S. Mohr, A. W. Holleitner, D. B. Bucher, A. V Stier, J. J. Finley, Unveiling the zero-phonon line of the boron vacancy center by cavity-enhanced emission. *Nano Lett.* **22**, 5137–5142 (2022).
35. S. A. Manuilov, C. H. Du, R. Adur, H. L. Wang, V. P. Bhallamudi, F. Y. Yang, P. C. Hammel, Spin pumping from spinwaves in thin film YIG. *Appl. Phys. Lett.* **107**, 042405 (2015).
36. C. W. Sandweg, Y. Kajiwara, A. V. Chumak, A. A. Serga, V. I. Vasyuchka, M. B. Jungfleisch, E. Saitoh, B. Hillebrands, Spin pumping by parametrically excited exchange magnons. *Phys. Rev. Lett.* **106**, 216601 (2011).
37. E. Lee-Wong, R. Xue, F. Ye, A. Kreisel, T. van der Sar, A. Yacoby, C. R. Du, Nanoscale detection of magnon excitations with variable wavevectors through a quantum spin sensor. *Nano Lett.* **20**, 3284–3290 (2020).
38. B. A. Kalinikos, A. N. Slavin, Theory of dipole-exchange spin wave spectrum for ferromagnetic films with mixed exchange boundary conditions. *J. Phys. C Solid State Phys.* **19**, 7013–7033 (1986).
39. A. Kreisel, F. Sauli, L. Bartosch, P. Kopietz, Microscopic spin-wave theory for yttrium-iron garnet films. *Eur. Phys. J. B* **71**, 59–68 (2009).
40. S. O. Demokritov, V. E. Demidov, O. Dzyapko, G. A. Melkov, A. A. Serga, B. Hillebrands, A. N. Slavin, Bose–Einstein condensation of quasi-equilibrium magnons at room temperature under pumping. *Nature* **443**, 430–433 (2006).
41. C. L. Jermain, S. V. Aradhya, N. D. Reynolds, R. A. Buhrman, J. T. Brangham, M. R. Page, P. C. Hammel, F. Y. Yang, D. C. Ralph, Increased low-temperature damping in yttrium iron garnet thin films. *Phys. Rev. B* **95**, 174411 (2017).
42. B. Flebus, P. Upadhyaya, R. A. Duine, Y. Tserkovnyak, Local thermomagnonic torques in two-fluid spin dynamics. Local thermomagnonic torques in two-fluid spin dynamics. *Phys. Rev. B* **94**, 214428 (2016).
43. S. Liu, R. He, L. Xue, J. Li, B. Liu, J. H. Edgar, Single crystal growth of millimeter-sized monoisotopic hexagonal boron nitride. *Chem. Mater.* **30**, 6222–6225 (2018).
44. R. Feynman, R. Leighton, M. Sands, *The Feynman Lectures on Physics* vol. 2 (Addison-Wesley Publishing Company, 1965).

**Acknowledgements**: **Funding:** J. Z. and C. R. D. acknowledged the support from U. S. National Science Foundation (NSF) under award No. DMR-2342569. H. L., M. H., H. W., and C. R. D. were supported by the Air Force Office of Scientific Research under award No. FA9550-20-1-0319 and its Young Investigator Program under award No. FA9550-21-1-0125. C. R. D. also acknowledges the support from the Office of Naval Research (ONR) under grant No. N00014-23-1-2146. F. A. and Z. J. were supported by NASA-REVEALS SSERVI (CAN No. NNA17BF68A) and NASA-CLEVER SSERVI (CAN No. 80NSSC23M0229).

**Competing interests**
The authors declare no competing interests.

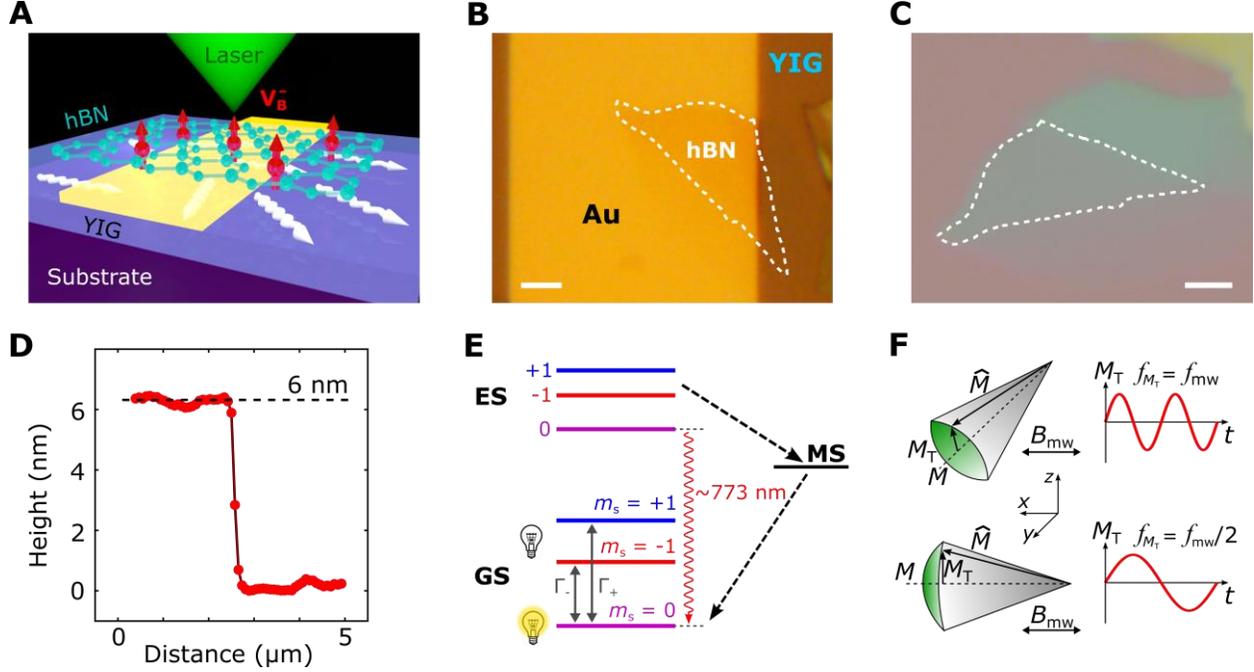

**Figure 1. Measurement platform and mechanism.** (**A**) Schematic of a hexagonal boron nitride (hBN) nanoflake transferred onto an Au microwave stripline patterned on magnetic insulator $Y_3Fe_5O_{12}$ (YIG) for quantum sensing measurements. (**B**)-(**C**) Optical microscope images of a prepared hBN-YIG device and the constituent hBN flake. The surveyed hBN flake area is outlined with white dashed lines, and the scale bar is 5 μm. (**D**) One-dimensional atomic force microscopy scan from which the thickness of the surveyed hBN flake area is characterized to be ~6 nm. (**E**) Energy level diagram of a $V_B^-$ spin defect and the nonradiative decay processes between the excited state (ES), metastable state (MS), and ground state (GS). $\Gamma_+$ and $\Gamma_-$ characterize the spin relaxation rates between $m_s = 0$ and $m_s = \pm 1$ states. (**F**) Schematic illustration of ferromagnetic resonance (FMR) and parametric spin excitations. The microwave magnetic field $B_{mw}$ is applied along the *x*-axis direction. The static YIG magnetization $M$ is perpendicular and parallel with $B_{mw}$ under the FMR and parametric pumping conditions, respectively. The transverse magnetization component $M_T$ oscillates at a frequency $f_{mw}$ and $f_{mw}/2$ under the FMR and parametric spin excitation geometries.

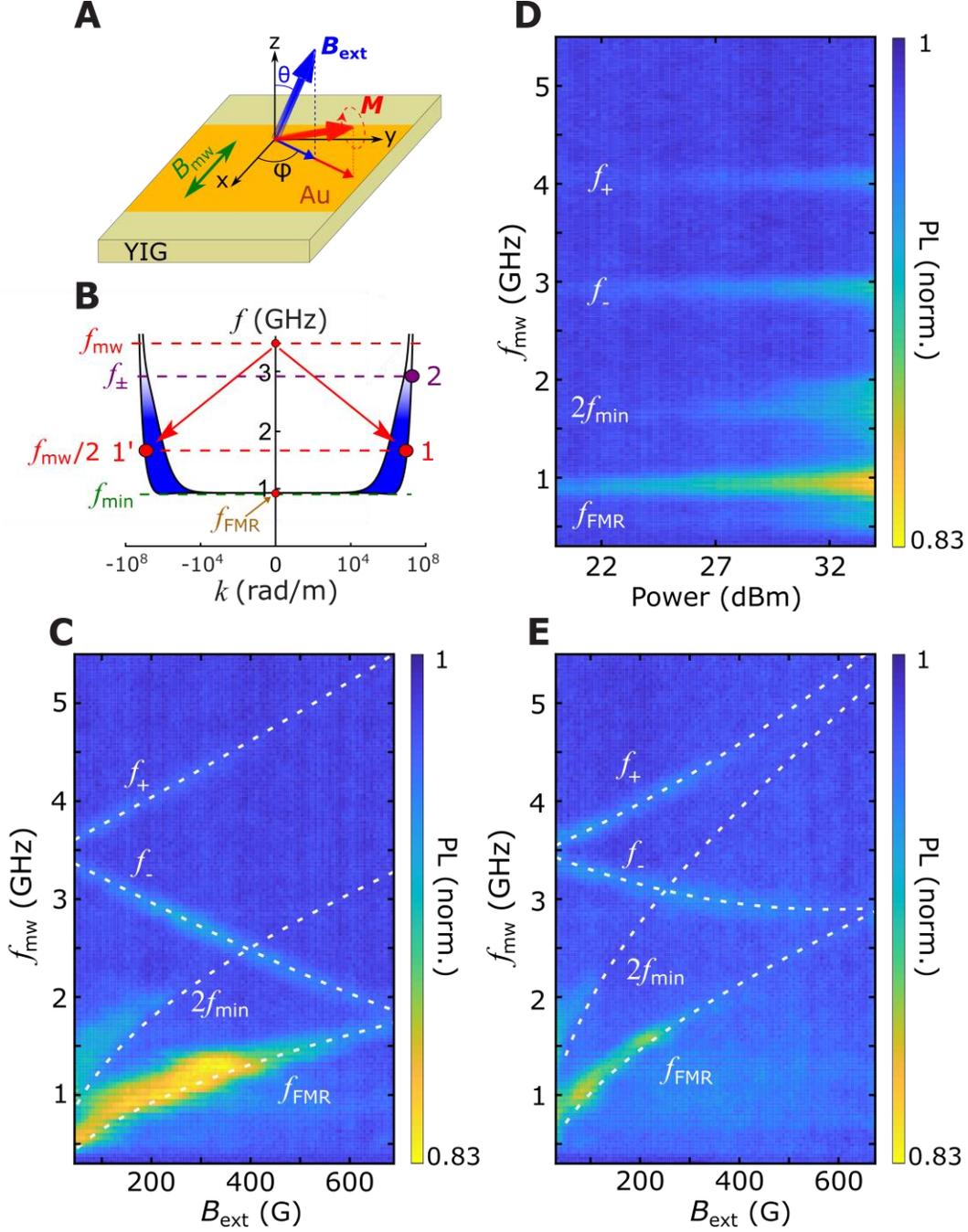

**Figure 2. Sensing Y$_3$Fe$_5$O$_{12}$ (YIG) spin waves using $V_B^-$ spin defects in 6-nm-thick hexagonal boron nitride.** (**A**) A coordinate system to illustrate the configuration of external magnetic field $B_{ext}$, YIG magnetization $M$, and local microwave magnetic field $B_{mw}$ in our measurements. (**B**) Sketch of the magnon band structure of a 100-nm-thick YIG thin film and illustration of the ferromagnetic resonance (FMR) and parametric spin excitations. The upper and lower branch of the dispersion curves correspond to the magnetostatic surface spin wave mode and the backward volume spin wave mode, respectively. The energy dependent (1/$E$) YIG magnon density is indicated by the fading color filled in the area between the two dispersion curves. The $V_B^-$ electron

spin resonance (ESR) frequency $f_\pm$ (purple color), FMR frequency $f_{FMR}$ (brown color), microwave drive frequency $f_{mw}$ (red color), half of the microwave drive frequency $f_{mw}/2$ (red color), and the magnon band minimum energy $f_{min}$ (green color) are marked in the picture of the YIG magnon band structure. (**C**) Normalized photoluminescence (PL) of $V_B^-$ spin defects measured as a function of external magnetic field $B_{ext}$ and microwave frequency $f_{mw}$. $B_{ext}$ is applied in the direction of $\theta = 16$ degrees and $\varphi = 39$ degrees. (**D**) A linecut at $B_{ext} = 204$ Oe of the optically detected magnetic resonance (ODMR) map measured over a broad range of the input microwave power from 20 dBm to 34 dBm. (**E**) Normalized PL map of $V_B^-$ spin defects measured when the $B_{ext}$ is 42 degrees tilted away from the out-of-plane direction. The in-plane azimuthal angle $\varphi$ of $B_{ext}$ is fixed at 39 degrees in this measurement. White dashed lines shown in Figs. 2**C** and 2**E** represent the calculated field dependent $f_{FMR}$, $2f_{min}$, and $V_B^-$ ESR curves.

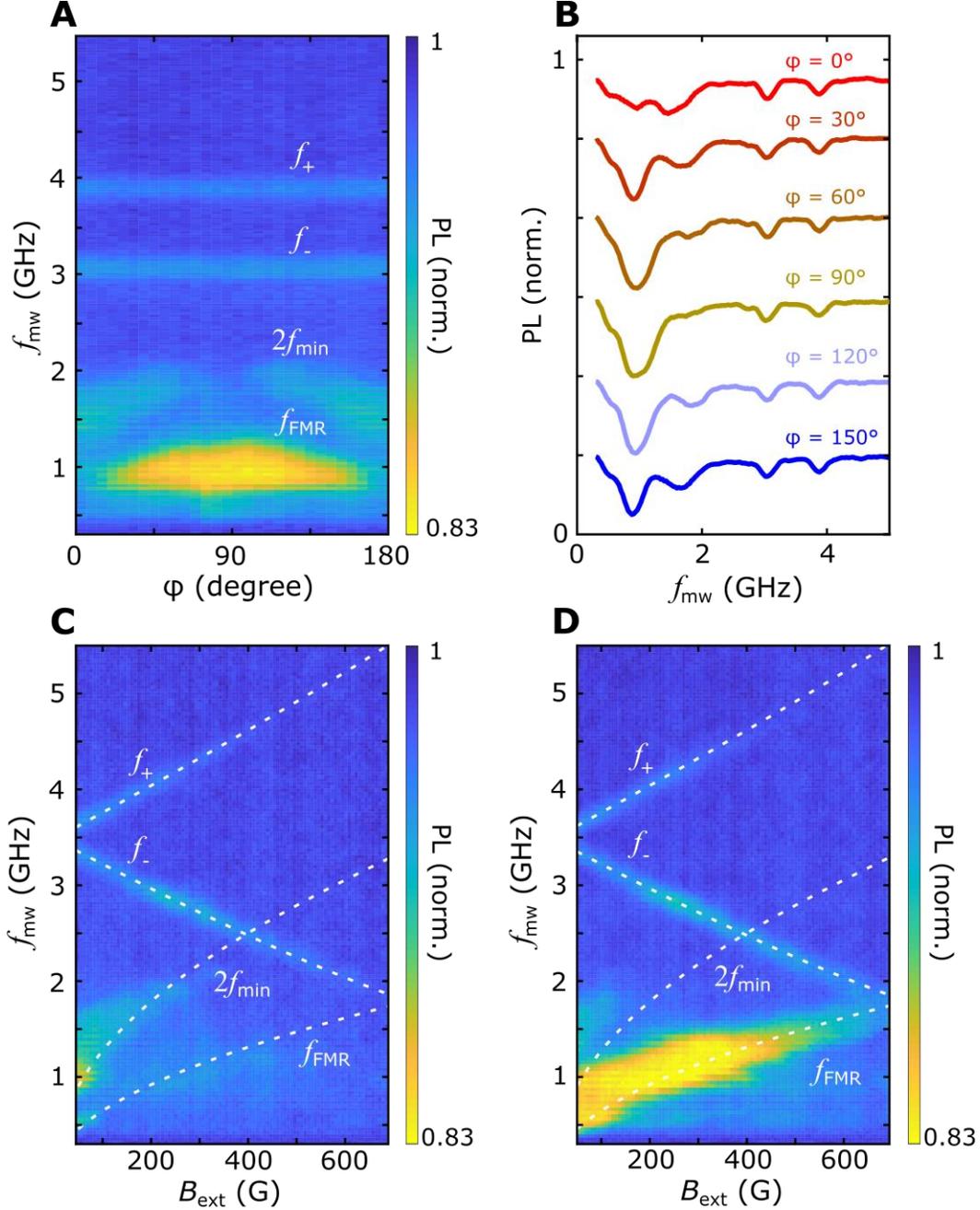

**Figure 3. Optically detected magnetic resonance (ODMR) spin wave spectra measured by a 6-nm-thick hexagonal boron nitride flake at different azimuthal angles of external magnetic field.** (**A**) Normalized photoluminescence (PL) of $V_B^-$ spin defects measured as a function of microwave frequency $f_{mw}$ and the in-plane azimuthal angle $\varphi$ of static magnetic field $B_{ext}$. The magnitude and polar angle $\theta$ of $B_{ext}$ is fixed at 150 Oe and 16 degrees for the measurement. (**B**) A series of linecut spectra at different azimuthal angle $\varphi$ of the ODMR map. The curves are offset for visual clarity. (**C**)-(**D**) Normalized PL maps of $V_B^-$ spin defects measured as a function of external magnetic field $B_{ext}$ and microwave frequency $f_{mw}$. $B_{ext}$ is applied along the direction of azimuthal angle $\varphi = 0$ degree (**C**) and $\varphi = 90$ degrees (**D**), respectively, and the polar angle $\theta$ is

set to be 16 degrees in these measurements. The white dashed lines represent the calculated field dependent $f_{\text{FMR}}$, $2f_{\text{min}}$, and $V_B^-$ electron spin resonance curves.

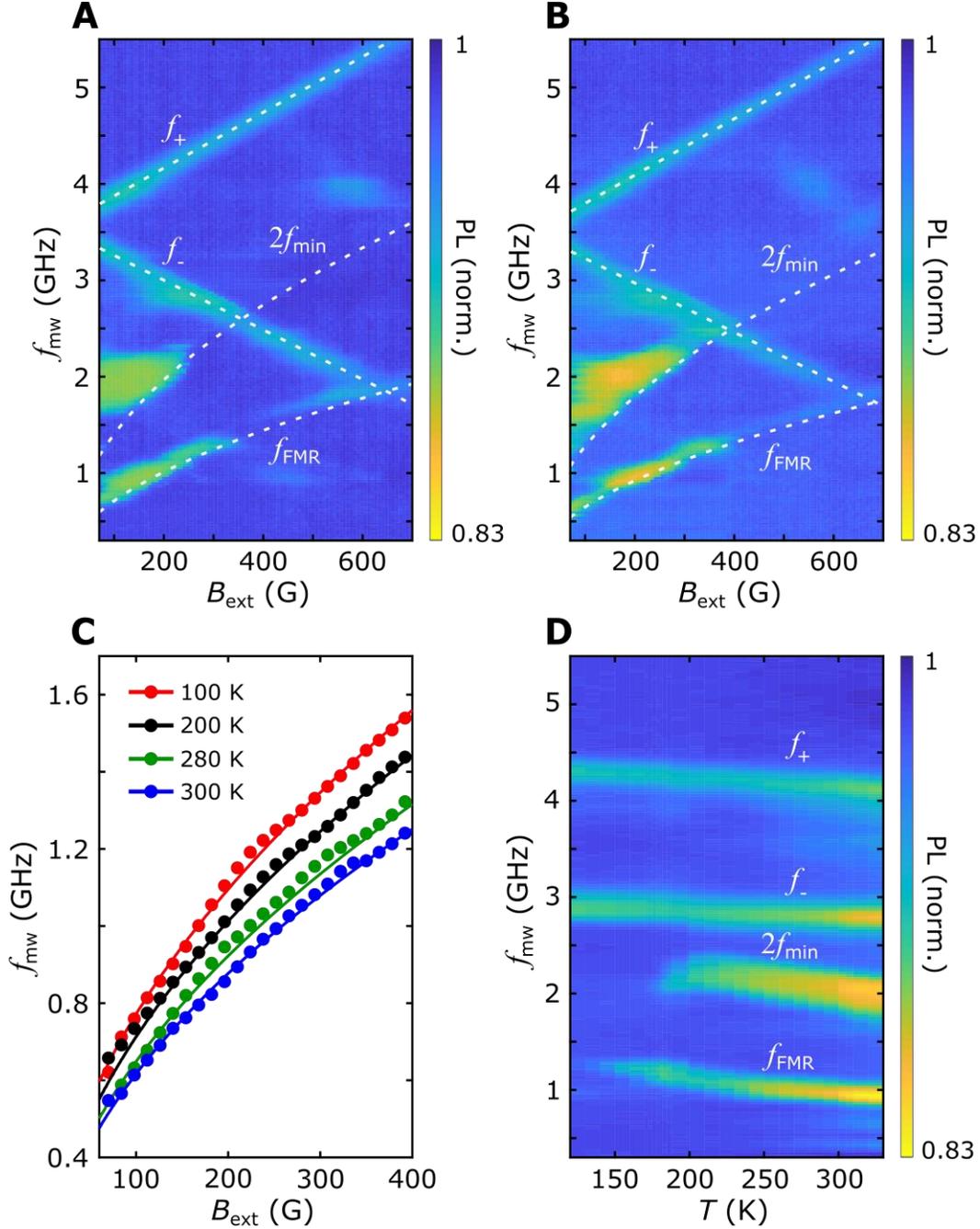

**Figure 4. Temperature dependence of optically detected magnetic resonance (ODMR) measurements of $Y_3Fe_5O_{12}$ (YIG) spin waves by a 76-nm-thick hexagonal boron nitride flake.** (**A**)-(**B**) Normalized photoluminescence maps of $V_B^-$ spin defects measured at 200 K (**A**) and 280 K (**B**). The external magnetic field $B_{ext}$ is 16 degrees tilted away from the out-of-plane direction and its in-plane azimuthal angle $\varphi$ is 15 degrees. The white dashed lines represent the calculated field dependent $f_{FMR}$, $2f_{min}$, and $V_B^-$ electron spin resonance (ESR) curves. (**C**) Magnetic field dependence of the ferromagnetic resonance (FMR) frequency $f_{FMR}$ of the 100-nm-thick YIG film measured under different temperatures. The lines and dots plotted represent theoretical calculations

and our experimental data, respectively. (**D**) A linecut at $B_{ext}$ = 225 Oe of the ODMR map measured from 100 K to 300 K. The input microwave power is fixed at 36 dBm.